\documentclass[sn-mathphys-num]{sn-jnl}

\usepackage{graphicx}%
\usepackage{multirow}%
\usepackage{amsmath,amssymb,amsfonts}%
\usepackage{amsthm}%
\usepackage{mathrsfs}%
\usepackage[title]{appendix}%
\usepackage{xcolor}%
\usepackage{textcomp}%
\usepackage{manyfoot}%
\usepackage{booktabs}%
\usepackage{algorithm}%
\usepackage{algorithmicx}%
\usepackage{algpseudocode}%
\usepackage{listings}%
\usepackage{comment}
\usepackage{hyperref}  
\usepackage{lscape}
\begin{document}

\title[Article Title]{Effect of modeling subject-specific cortical folds on brain injury risk prediction under blunt impact loading}

\author[1]{\fnm{Anu} \sur{Tripathi}}\email{tripathia@rmu.edu}

\author[2]{\fnm{Alison} \sur{Brooks}}\email{brooks@ortho.wisc.edu}

\author[3]{\fnm{Traci} \sur{Snedden}}\email{traci.snedden@cuanschutz.edu}

\author[4]{\fnm{Peter} \sur{Ferrazzano}}\email{ferrazzano@pediatrics.wisc.edu}

\author[5]{\fnm{Christian} \sur{Franck}}\email{cfranck@wisc.edu}

\author*[1]{\fnm{Rika Wright} \sur{Carlsen}}\email{carlsen@rmu.edu}

\affil[1]{\orgdiv{Department of Engineering}, \orgname{Robert Morris University}, \orgaddress{\city{Moon Township}, \state{PA}, \country{USA}}}

\affil[2]{\orgdiv{Department of Orthopedics and Rehabilitation}, \orgname{University of Wisconsin--Madison}, \orgaddress{\city{Madison}, \state{WI}, \country{USA}}}

\affil[3]{\orgdiv{College of Nursing}, \orgname{University of Colorado Anschutz Medical Campus}, \orgaddress{\city{Aurora}, \state{CO}, \country{USA}}}

\affil[4]{\orgdiv{Waisman Center}, \orgname{University of Wisconsin--Madison}, \orgaddress{\city{Madison}, \state{WI}, \country{USA}}}

\affil[5]{\orgdiv{Department of Mechanical Engineering}, \orgname{University of Wisconsin--Madison}, \orgaddress{\city{Madison}, \state{WI}, \country{USA}}}

\abstract{

\textbf{Purpose} 
Computational head models are essential tools for predicting the risk of mild traumatic brain injury (mTBI) under different activities and across populations. However, different computational models incorporate varied levels of anatomical details, such as cortical folds. 
In this study, we aim to determine the effect of modeling cortical folds on mTBI risk assessment. \\
\textbf{Methods} 
We compared gyrencephalic (with cortical folds) and lissencephalic (without cortical folds) FE models of 18 subjects aged 9 - 18 years under a rotational head acceleration event. A rotational acceleration of 10 krad/s$^2$ and 10 ms duration was simulated about each principal head axis. 
We analyzed different tissue-level mTBI injury metrics, including maximum principal strain (MPS95), maximum principal strain rate (MPSR95), and cumulative strain damage measure (CSDM15), for the whole brain and for specific regions of interest (ROIs).  \\
\textbf{Results} 
The inclusion of cortical folds consistently yielded higher injury metrics across all individuals and rotational directions, with a bias (mean $\pm$ std. dev. relative to maximum lissencephalic values) of 
$-21.7 \pm 9.1 \%$ in MPS95, $-17.1\pm7.6\% $ in MPSR95, and $-14.4\pm11.3\% $ in CSDM15. 
Modeling cortical folds significantly affected the spatial strain distributions, with the DICE similarity coefficient on peak MPS ranging between $0.07-0.43$ and on CSDM15 ranging between $0.42-0.70$; and increasing the peak injury metrics even in the geometrically unaltered regions of interest, such as the corpus callosum, cerebellum, and brain stem, by up to $\sim$50\%.
\textbf{Conclusions} 
The study finds that the inclusion of cortical folds significantly alters the pattern of deformation in the brain, thereby affecting the mTBI risk predictions head rotations.
}

\keywords{Mild traumatic brain injury, Finite element head models, Subject specific modeling, Cortical folds}

\maketitle
\section{Introduction}\label{Introduction}

Mild traumatic brain injury (mTBI), including concussion, is a highly prevalent and heterogeneous condition, affecting different individuals differently \cite{langdon2023heterogeneity}. Computational head models have been used to study the mTBI risk of various activities in different populations 
\cite{liu2022influence,filben2021header,takhounts2008investigation,nakarmi2024role,brooks2021purposeful,zimmerman2021player}. These head models vary significantly in terms of the incorporated level of anatomical detail, the individual anatomical variations, and the choice of finite element (FE) mesh type or material models.  
Cortical folds, which consist of the gyri and sulci in the brain's cerebrum, are one such anatomical detail. 
While some studies have used models with cortical folds to study the risk of mTBI \cite{reynier2022quantifying,zimmerman2021player,miller2021brain,filben2021header}, others use FE models without cortical folds \cite{cecchi2021identifying,brooks2021purposeful,liu2022influence}.
However, the effect of incorporating cortical folds into FE head models on mTBI risk assessment is not well understood. 
In this paper, we aim to study the effects of modeling several subject-specific cortical folds on predicting the mTBI risk under head rotational accelerations.

Previous studies have investigated the role of cortical folds on tissue level mTBI injury metrics, such as strains \cite{song2015finite,ho2009can,saez2020topological,mazurkiewicz2021impact,fagan2020simulation} and stresses \cite{saez2020topological,fagan2020simulation,song2015finite} in the brain tissue. These computational \cite{song2015finite,ho2009can,saez2020topological,fagan2020simulation} and experimental \cite{mazurkiewicz2021impact} studies compared the brain strains and stresses developed between a gyrencephalic (with cortical folds) and a lissencephalic (smooth cortical surface without folds) model, but resulted in conflicting findings. Some computational studies based on 3D \cite{ho2009can} and 2D sagittal models \cite{song2015finite} of the human head, and 3D head models of multiple species \cite{saez2020topological} reported higher strain and stress in the lissencephalic models. On the other hand, an experimental study on 2D coronal surrogates of porcine brain \cite{mazurkiewicz2021impact} and a computational study on 2D-axial human brain model \cite{fagan2020simulation} found higher strains in gyrencephalic models. These discrepancies could stem from differences in the impact loading conditions, tissue material properties, and the geometrical variations in the cortical folds modeled across the studies. 


Some computational studies have explored the role of geometrical variations in cortical folds on the deformation in the brain tissue \cite{cloots2008biomechanics,bakhtiarydavijani2021mesoscale,saboori2012effect,he2022mesoscale}. These studies are based on 2D plane-strain mesoscale models of a square region incorporating approximately a couple of gyri with simplified geometry. 
These studies also reported conflicting trends, with most of the studies finding higher strains in gyrencephalic models \cite{bakhtiarydavijani2021mesoscale,cloots2008biomechanics,he2022mesoscale}, but one study reporting higher strains in the lissencephalic model \cite{saboori2012effect}. 
The models used in these studies varied in terms of the applied boundary conditions to approximate an impact loading, with some studies applying uniform compression on different sides of the mesoscale model \cite{he2022mesoscale,saboori2012effect,bakhtiarydavijani2021mesoscale}, and others applying non-uniform shear acceleration field obtained from a full head simulation \cite{cloots2008biomechanics}.
The studies also had different tissue mechanical properties.
Also, since the variations in the cortical folds were not based on actual human brain anatomy, the practical and statistical significance of incorporating cortical folds in subject-specific FE head models for mTBI risk assessment remains unanswered, and the computational studies on mTBI continue to use both lissencephalic and gyrencephalic models to understand mTBI risk in practical scenarios \cite{liu2022influence,filben2021header}.

In this study, we investigate the role of cortical folds on mTBI risk, while accounting for cortical fold variations across different individuals and the directions of loading. 
We develop subject-specific gyrencephalic and lissencephalic models of 18 subjects and compare their mTBI response under head rotations about the three principal anatomical axes.
The results of this study will provide the importance of incorporating cortical fold details in FE head models used for mTBI risk prediction.

The paper is organized as follows. Section \ref{sec:method} describes the methods and workflows used in developing the subject-specific models (gyrencephalic and lissencephalic) from their medical images and the details of the finite element simulations. Section 3 provides the results on different mTBI metrics from the computational study, and Section 4 discusses the implications of these results on our understanding and future steps.

\section{Materials and Methods}\label{sec:method}
In this section, we describe the development of gyrencephalic (with cortical folds) and lissencephalic (without cortical folds) finite element (FE) head models (Section \ref{sec:FEMdevelopment}), followed by the details of FE simulation (Section \ref{sec:FESim}), injury metrics (Section \ref{sec:InjuryMetrics}), and data analyses (Section \ref{sec:DataAnalyses}) to detail the differences between the mTBI risk assessment using the two model types.

\subsection{Subject-specific Head Modeling}
\label{sec:FEMdevelopment}
We developed the FE head models of 18 individuals (8 males, 10 females; 9--18 years old) from their magnetic resonance imaging (MRI) scans. The number of subjects by age and sex in each group is provided in Table \ref{AgeSex}. 
Subject-specific gyrencephalic and lissencephalic models were generated directly from the medical images. 

\begin{table}[h!]
\centering
\begin{tabular}{c c c}
\hline
\textbf & \textbf{9 -- 13 Years} & \textbf{16 -- 18 Years} \\
\hline
Male   & 5 & 3 \\
\hline
Female & 6 & 4 \\
\hline
\end{tabular}
\caption{Number of subjects by age and sex}
\end{table}\label{AgeSex}

\subsubsection{Medical Image Segmentation}
\paragraph{Gyrencephalic Models}
The first step in generating a medical image-based FE model is medical image segmentation, which delineates the boundary of different parts of the model. 
The medical images of each individual consisted of structural MRI (T1-w) and diffusion tensor imaging (DTI) (Fig. \ref{Fig:workflow}a).

The T1-structural MRIs were 
first pre-processed to correct for Gibbs-ringing and bias-field artifact using open source software MRtrix \cite{tournier2019mrtrix3} and ANTS \cite{tustison2014large}, respectively. All MRI scans were registered rigidly to the MNI152 atlas using 3D slicer (http://www.slicer.org) \cite{fedorov20123d}. 
The segmentation of the brain tissue into the cortical gray matter, white matter, brain stem, and deep brain regions was obtained using the FreeSurfer software package \cite{fischl2012freesurfer}. The resulting gray matter incorporates cortical folds, or gyri and sulci, in 68 parcellations (Fig. \ref{Fig:workflow}b), providing a gyrencephalic head model segmentation. 
Automated brain segmentation on the enhanced T1 MRIs using the FSL software \cite{smith2004advances} provided the skull segment. 
The additional 1 mm thick meninges (dura, falx, and tentorium) segment was added manually after combining the FreeSurfer and FSL segmentations using 3D Slicer. The sub-arachnoid cerebrospinal fluid (CSF) was filled in the space between the meninges and the brain segments. A one mm layer of CSF was added around the meninges to ensured that brain tissue was surrounded by the CSF everywhere. This workflow provided the gyrencephalic (with cortical folds) model segmentations.


\paragraph{Lissencephalic Models}
The lissencephalic models were generated by modifying the detailed gyrencephalic model segmentations. First, the heterogeneous sub-arachnoid CSF in the gyrencephalic segmentation was replaced by a uniform 1 mm thick layer of CSF lining the inner surface of the meninges. The cortical gray and white matter matter with gyri and sulci in the gyrencephalic segmentation were replaced by a uniform 4 mm thick gray matter layer inside the new uniform CSF layer segment. The remaining space between the gray matter and the deep brain structures was filled with white matter, to provide the lissencephalic model segmentations (Fig. \ref{Fig:workflow}b).

\subsubsection{Finite Element Meshing}

The FE mesh was generated directly from the segmentation by converting the segmented image voxels (1 mm $\times$ 1 mm $\times$ 1 mm) directly to hexahedral elements using a custom MATLAB script (Mathworks Inc.) (Fig. \ref{Fig:workflow}c). The voxel meshing allowed fast mesh generation ($\sim15$ minutes on 1 CPU) while retaining fine subject-specific details, such as cortical folds. While the interfaces between two materials were jagged and non-smooth, a recent study found that the voxel and conformal mesh (with smooth interfaces) result in identical response when comparing strain based injury metrics, such as the 99th percentile strain or the volume fraction of brain with strains over certain thresholds \cite{zhou2025surface}. A number of FE head models that incorporate cortical folds use voxel mesh \cite{ho2008generation,giudice2021calibration,miller2016anatomically}. A reduced integration scheme with hourglass control was implemented (element C3D8R) to prevent volumetric locking behavior given the nearly incompressible material behavior of the brain tissues. 


\subsubsection{Material Modeling}
The brain tissue was modeled as an anisotropic visco-hyperelastic material with the axonal tract orientation from the DTI scans. The time dependence is captured using a first-order Prony series. The bulk response of the CSF was modeled using the Mie Gruneisen equation of state, and the shear response as Newtonian viscous flow. The meninges were modeled as a linear elastic solid, and skull as a rigid solid. The material model definitions of the brain tissue, CSF, and the meninges were based on a previously published FE head model \cite{nakarmi2024role,wright2013multiscale}. The evaluation of the brain mechanical response one of the subject-specific FE models used in this study against experimental data is presented in the Supplementary Sections S2 and S3. 

\subsection{Finite Element Simulations}
\label{sec:FESim}
\begin{figure}[htbp]
    \centering
    \includegraphics[width=0.9\textwidth]{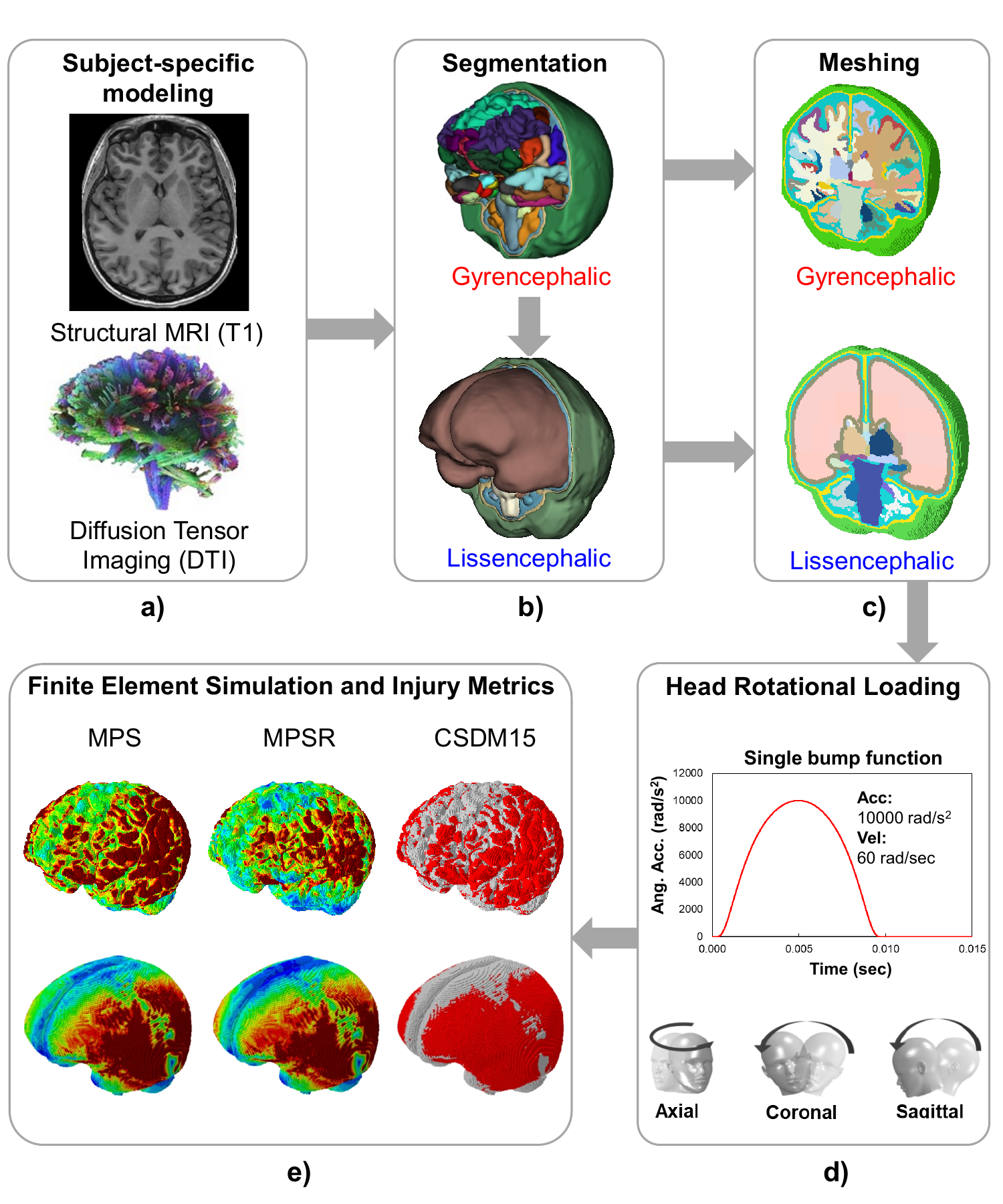}
    \caption{Workflow for studying the effect of modeling cortical folds on mTBI injury metrics: a) Subject-specific magnetic resonance imaging (MRI) scans, b) Medical image segmentation to create gyrencephalic and lissencephalic head models,
    c) Voxel meshing, d) Idealized head rotation acceleration profile applied about the three cardinal anatomical planes, and
    e) Post-processing to obtain injury metrics}
    \label{Fig:workflow}
\end{figure}

Each finite element head model was used to simulate three head acceleration events, where a half-sinusoidal rotational acceleration pulse was applied to the skull about a principal anatomical axis (axial, sagittal, and coronal) \cite{carlsen2021quantitative}. The peak acceleration was selected to be 10 krad/s$^2$ for a duration of 10 ms, resulting in a peak angular velocity of 60 rad/sec (Fig \ref{Fig:workflow}d). The loading magnitude was selected based on measured data in sports such as mixed martial arts and American football, to represent an event with a high probability of concussion \cite{tiernan2020concussion, laksari2020multi}.
The simulations were conducted using an explicit time integration scheme in Abaqus FE software (Simulia, Dassault).%

\subsection{Injury Metrics}
\label{sec:InjuryMetrics}

To quantify the effect of cortical folds on mTBI risk assessment, we compare different tissue-level injury metrics obtained from the gyrencephalic and lissencephalic model simulations for each subject. Previous studies have used the strain and strain rate in the brain tissue from FE simulations with mTBI risk \cite{gabler2016assessment,wu2021head,margulies1992proposed,zhang2024objective}. 
The FE simulations in Abaqus/Explicit provided the maximum principal logarithmic strain (MPS) and maximum principal logarithmic strain rate (MPSR) at each material (element) and time point \cite{carlsen2021quantitative}.

An element was flagged as damaged after the MPS at its location exceeded 0.15, based on experimental neuronal injury thresholds \cite{hajiaghamemar2020head,bain2000tissue,bar2016strain,estrada2021neural,gonzalez2024cortical,cullen2007strain}. A cumulative strain damage measure (CSDM15) was calculated as the fraction of the damaged elements' volume to the total brain tissue volume. We also calculated the cumulative strain rate damage measure (CSRDM50) based on elements exceeding MPSR of 50 s$^{-1}$ for a representative case. 

Since the peak MPS at a given material point occurs at different times in the two models, we calculated the time-invariant peak strain distribution for a given simulation as the peak MPS experienced by an element over the whole duration of the head acceleration simulation (30 ms) \cite{upadhyay2024effect}. We refer to this time-invariant peak MPS distribution as just `the MPS distribution' in the following sections, unless otherwise stated. This allows the comparison of peak strain experienced at the same anatomical locations between the two models, i.e. the similarity of strain distribution. 
We also calculated the time-invariant peak MPSR distributions for a representative case.

To study the effect of cortical folds on the overall global response of the brain during a head acceleration event, we compare the highest strains and strain rates developed anywhere within the brain tissue at any point during the simulation in the gyrencephalic and lissencephalic models of a given subject. 
The FE simulations were post-processed to provide the 95th percentile of the MPS (MPS95) and MPSR (MPSR95) (instead of the 100th percentile, which is commonly avoided in complex biological FE models to avoid numerical artifacts).

\subsection{Data Analysis}
\label{sec:DataAnalyses}

We use Sørensen DICE coefficient (DICE) to compare the similarity of the spatial distributions in the gyrencephalic and lissencephalic models. The DICE coefficient was obtained for the MPS and CSDM distributions, as defined below:
\begin{equation}
    \text{DICE} = \frac{|G \cap L|}{|G|}
\end{equation}

\noindent where $G$ and $L$ are the element sets in the gyrencephalic and the lissencephalic models, 
$|G \cap L|$ is the intersection of the two sets, i.e. elements at corresponding spatial locations in both models that have the same MPS (or CSDM) values. We consider an element to have the same values in the two models if the absolute difference is less than 0.025. The DICE value is sensitive to the threshold, and we selected the value of 0.025 to be consistent with the contour plots showing MPS distribution in the results section (Section \ref{Results}).   
A higher value of DICE indicates a better spatial agreement between the two models, with 1 indicating all elements have the same values and 0 being no elements have the same values. The DICE coefficient for MPS represents the similarity over the whole range of strain values, whereas the DICE coefficient for the CSDM represents the similarity in damaged elements in the two models.

We calculate the means and standard deviations of the peak values of the injury metrics (MPS95, MPSR95, and CSDM15) from all subjects to compare the gyrencephalic and lissencephalic models. We perform this comparison across the whole brain, and specific anatomical regions of interest (ROI), including cerebellum, brain, corpus callosum, cerebral gray matter, cerebral white matter, and deep brain region comprising the thalamus, hippocampus, amygdala, putamen, pallidum, and hypothalamus. 
We performed the Bland-Altman test to obtain the bias between the gyrencephalic and lissencephalic models for peak MPS95, peak MPSR95, and CSDM15. We normalize the bias by the maximum lissencephalic result for the case to understand the practical significance of the bias. 
We performed a statistical significance test using paired t-test, and $p<0.005$ is considered significant.

\section{Results}
\label{Results}

We developed subject-specific gyrencephalic and lissencephalic FE head models of 18 individuals from their MRI scans. Each model consisted between $\sim1.7-2.7 \times 10^6$ voxel hexahedral elements. We simulated a concussive blunt impact on each model through rigid body rotation of the skull (Fig \ref{Fig:workflow}d). The simulations were performed on Abaqus/Explicit and each simulation took 10-14 hours using 64 cores. In this section, we describe the tissue-level injury metrics (Section \ref{sec:InjuryMetrics}) obtained from these simulations to understand the effect of modeling cortical folds.

\subsection{Effect of modeling cortical folds on injury metrics: A representative case}

This section analyzes the spatial evolution of the strain and strain rates during axial head rotation in a representative subject (18F) to identify the key differences in trends between the gyrencephalic and lissencephalic models (Figure \ref{fig:rep_sub}). 
The spatial distributions are shown across an axial slice passing through the anterior and posterior horns of the lateral ventricles, parallel to the horizontal plane for the two models. 

\subsubsection{Brain tissue strain}\label{Diss:strain}

We found some similarities in the overall brain strain waves in the two models. 
The MPS distributions at different times show that the strain begins to increase at the outer surface of the brain (cortical surface) and travels to the center of the brain in both the gyrencephalic and lissencephalic models (Figure \ref{fig:rep_sub}a). 
We see higher strain concentration in the cerebral cortex in both models, although to different degrees (Figure \ref{fig:rep_sub}a).
Since a head acceleration event is modeled through providing the skull with a rigid body acceleration, a strain wave originates at the skull and travels inwards to the brain through the meninges and CSF. While the spherical convergence of the strain wave tends to increase the strain magnitude as the wave travels inwards, the high viscosity of the brain tissue causes dissipation of the strain magnitude as the wave travels inwards, and therefore results in higher strains on the cortical surface (Figure 2a-f) \cite{massouros2014strain,chen2010mri}. 

We observed differences between the two models in terms of the location of high-strain regions on the cortical surface. The lissencephalic model experiences higher strain on the lateral regions, which decreases gradually moving towards the falx cerebri along the cortical surface. Whereas, the high strain regions in the gyrencephalic model are non-uniform.
In the gyrencephalic model, the space between sulci is filled with CSF, which is significantly softer than the brain tissue material occupying the space in the lissencephalic model \cite{wright2013multiscale}. Therefore, modeling cortical folds directly impacts the effective stiffness of the skull-brain interface, which plays a significant role in the strains experienced by the brain tissue.
The thickness of the CSF layer varies across gyri, contributing to the heterogeneity of the high-strain region on the cortical surface. Also, the size and orientation of different sulci vary, resulting in different moments of inertia about the rotation axis, which can also affect the strain distributions. As the strain wave travels inwards, the geometric discontinuity from sulci (high curvature voids) causes higher strain at the tip of the sulci as compared to the adjacent tissue lying at the same depth from the surface of the gyri.

We observed further differences between the models as the strain wave reaches the tissue around the ventricles.
The strain localization in the gyrencephalic model near the ventricles is significantly higher than in the lissencephalic model, especially between the anterior horns of the lateral ventricles and the Sylvian sulcus. 
The lissencephalic model experiences strain localization due to the high-curvature voids formed by the anterior horns of the lateral ventricles.
In the gyrencephalic model, the incident strain wave in this region around the ventricles has a higher magnitude due to strain concentrations at the tip of the sulci filled with soft CSF. This strain is magnified further, similar to the lissencephalic model, due to the presence of high curvature voids (the anterior horns of the lateral ventricles). It should be noted that the high nonlinearity of the brain tissue results in large regions of strain concentration at the tip of the ventricles and sulci, which interact to result in greater strain concentration in this region compared to the lissencephalic model. Essentially, the region acts as a single-edge notch specimen in the lissencephalic model, and a double-edge notch specimen in the gyrencephalic model. 
The difference in the spatial MPS distribution is reflected in a low DICE similarity coefficient of 0.22 between the models. 
These differences also resulted in the peak MPS95 to be lower in the lissencephalic model (peak MPS95$_L =0.288$) than in the gyrencephalic model (peak MPS95$_G =0.345$). 


Overall, the lissencephalic model experienced high strains near the outer surface of the brain, whereas the gyrencephalic model experienced its highest strains near the ventricles and deeper regions of the brain in addition to the outer cortical surface (Figure \ref{fig:rep_sub}c), causing the time to peak MPS95 to be lower in the lissencephalic model ($t_L=10$ms) than in the gyrencephalic model ($t_G=12$ms). The CSDM15, which is the region above an injury threshold, shows a similar trend in the damaged regions (Figure \ref{fig:rep_sub}d). 
The spatial CSDM15 distribution has a higher DICE similarity coefficient of 0.67 between the models.

\subsubsection{Brain tissue strain rate}

\begin{landscape}
\begin{figure}[htbp]
    \centering
    \includegraphics[height=0.9\textwidth]{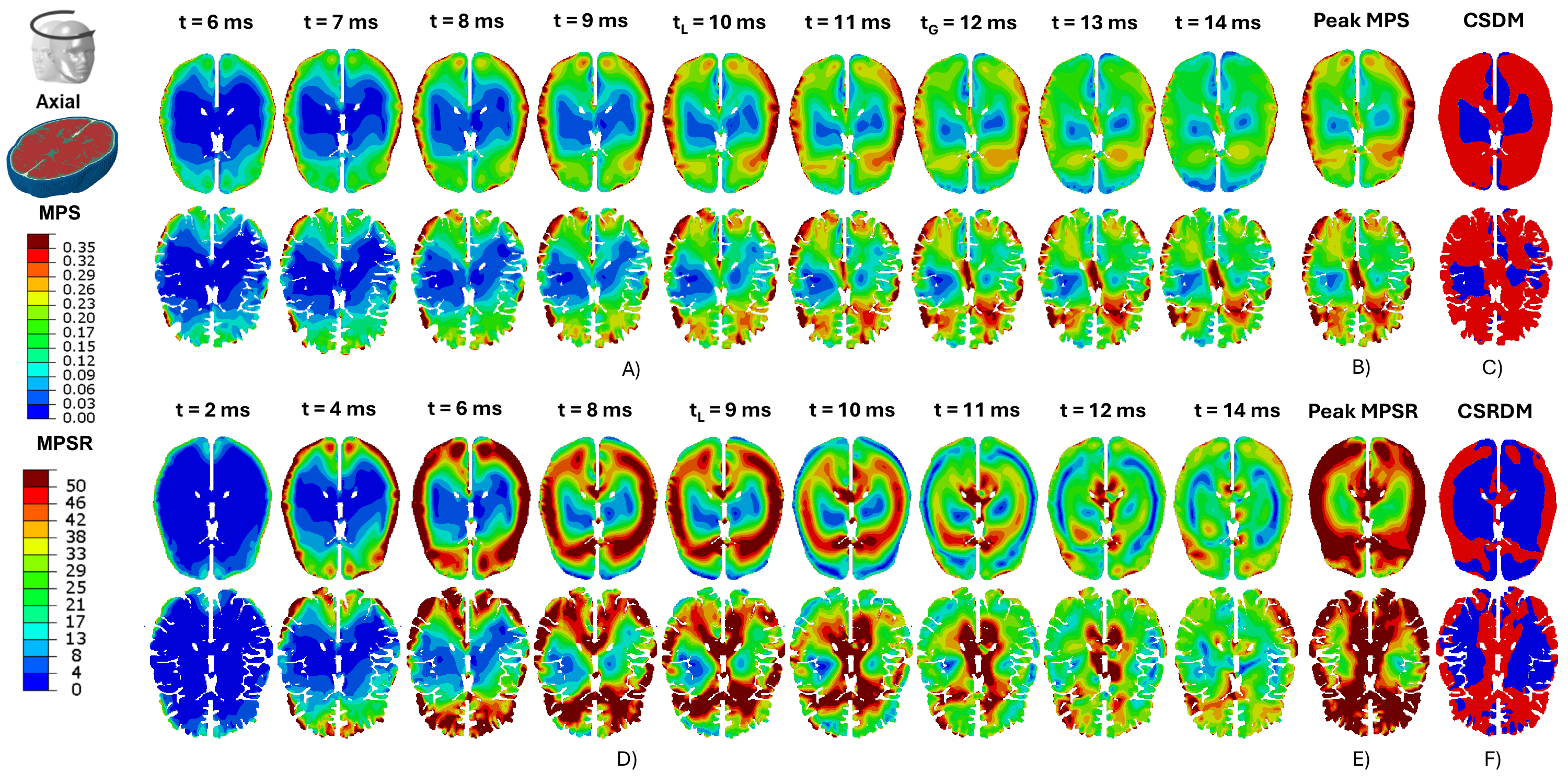}
    \caption{a) Maximum principal strain (MPS) contour plots of a representative subject at different times near MPS95 peak from the lissencephalic and the gyrencephalic models, b) the peak MPS (MPS) experienced by the element over the entire duration of the head acceleration event, c) damaged element shown in red after the MPS at its location exceeded 0.15.
    These plots highlight the spatial variations arising due to the cortical folds.}
    \label{fig:rep_sub}
\end{figure}
\end{landscape}

The MPSR distributions show similar trends as MPS while comparing the gyrencephalic and the lissencephalic models (Figure \ref{fig:rep_sub}d-f). 
The MPSR distributions show the strain rate wave propagating from the outer cortical surface to the center of the brain, while dissipating slowly (Figure \ref{fig:rep_sub}d). 
The lissencephalic model shows higher MPSR and CSRDM50 on the cortical surface than the gyrencephalic model distributions (Figure \ref{fig:rep_sub}e-f).
On the other hand, the highest MPSR and CSRDM50 in the gyrencephalic model are seen near the center of the brain around the ventricles, due to the interaction between the regions of strain concentrations at the boundary of voids created by the sulci and the ventricles, as explained in Section \ref{Diss:strain}.

Similarly, we observe lower peak MPSR95 and time to peak MPSR95 in the lissencephalic model (peak MPSR95$_L =$ 56.5 s$^{-1}$, $t_L=6$ ms) than in the gyrencephalic model (peak MPSR95$_G =$ 67.9 s$^{-1}$, $t_G=7$ ms).
Since the strain and strain rate differences follow similar trends, we only analyze the strain distribution differences across multiple subjects and loading directions in the following section. 

\subsection{Inter-subject variability in injury risk due to modeling cortical folds}

We analyze the injury metrics from the gyrencephalic and lissencephalic models for multiple subjects under different loading directions to understand the statistical significance of the contribution of cortical folds in mTBI risk assessment.

\subsection{Effect on spatial strain distributions}

\begin{landscape}
\begin{figure}[htbp]
    \centering
    \includegraphics[height=0.99\textwidth]{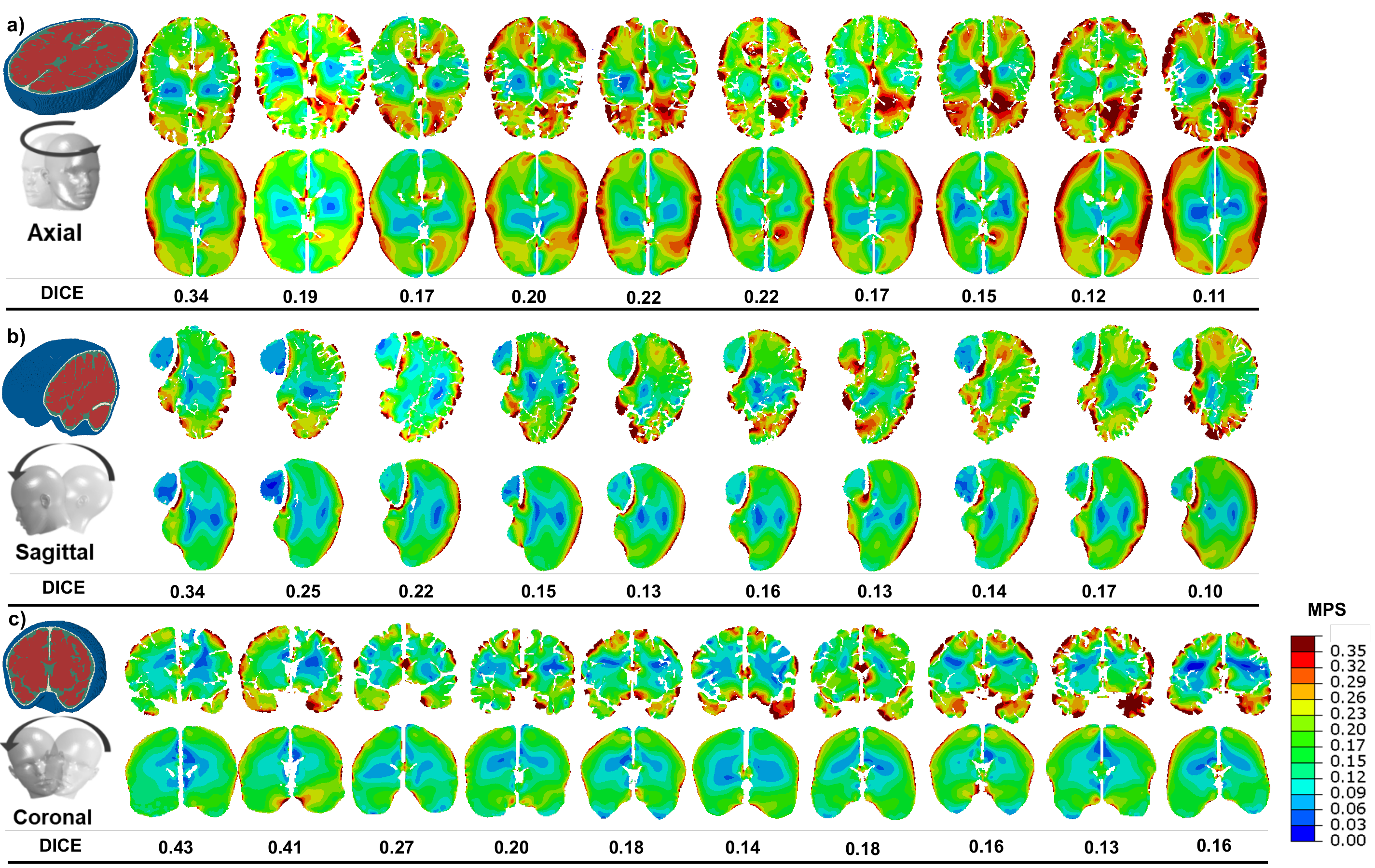}
    \caption{Peak maximum principal strain (MPS) contour plots of multiple representative subjects in the lissencephalic and the gyrencephalic models under rotational acceleration about each principal axis: a) axial, b) sagittal, and c) coronal. These MPS distributions capture the interaction of cortical folds with ventricles to result in different strain localization, especially around deeper sulci, such as the lateral (Sylvian) sulcus.
    }
    \label{fig:Strain_contours}
\end{figure}
\end{landscape}

Both the gyrencephalic and lissencephalic models of all subjects experienced higher strains on the outer cortical surface under all rotation directions. 
The high-strain regions on the outer cortical surface in lissencephalic models are present consistently along the smaller axis of the cross-section ellipse, where the radius of curvature is high, which allows the brain tissue to deform unconstrained under all loading directions. These high-strain regions on the cortical surface include i) the lateral regions away from the falx cerebri under the axial rotations (Figure \ref{fig:Strain_contours}a); ii) regions along the tentorium cerebeli, along the convex curvature regions between the frontal and the temporal lobe, and along the top cortical surface under the sagittal rotations (Figure \ref{fig:Strain_contours}b); and iii) on the lateral cortical surface away from the falx cerebri and the convex curvature region near the inferior cortical surface under the coronal rotations (Figure \ref{fig:Strain_contours}c). (These regions correspond to high-strain regions in solids with different cross-sections under rotation, Supplementary Section S5).
In the previous section, we saw that the skull-brain interface plays a significant role in governing the brain deformation, and the gyrencephalic model captures the heterogeneous skull-brain interface. Therefore, the individual variations in the cortical folds make the high-strain regions in the gyrencephalic model heterogeneous along the cortical surface as well as across different individuals. The CSF thickness between the brain and the skull also has a cushioning effect, dissipating the strain reaching the brain tissue, (Supplementary Section S4), also contributing to the hetergeneity in the brain tissue strain along the cortical surface.

In the inner regions of the brain, we observe more strain concentrations in the gyrencephalic model across all individuals and loading directions. 
Under axial rotations, the high strain concentration regions between the anterior horns of the lateral ventricles and the lateral (Sylvian) sulcus in gyrencephalic models become increasingly larger with increasing head size (moment of inertia, MoI about the axial axis) as compared to the lissencephalic models. 
Under sagittal rotations, the gyrencephalic models experience higher strains near the posterior horns of the lateral ventricles and the sucli in the frontal and temporal lobes. Similarly, under the coronal rotations as well, the gyrencephalic models show high strain regions on the outer cortical surface and around the Sylvian sulcus in the frontal and temporal lobes. The individual differences in the cortical fold geometry cause higher individual variations in the gyrencephalic model. 
In the previous section, we also saw that the distance between the ventricles and sulci affects the strain concentration in the gyrencephalic model, and consequently, the individual variations contribute to higher variations in the high-strain regions in the gyrencephalic models across individuals \cite{benko2021mechanical}. 
These differences across individuals are not pronounced in the lissencephalic models.

\begin{figure}[htbp]
    \centering
    \includegraphics[width=0.6\textwidth]{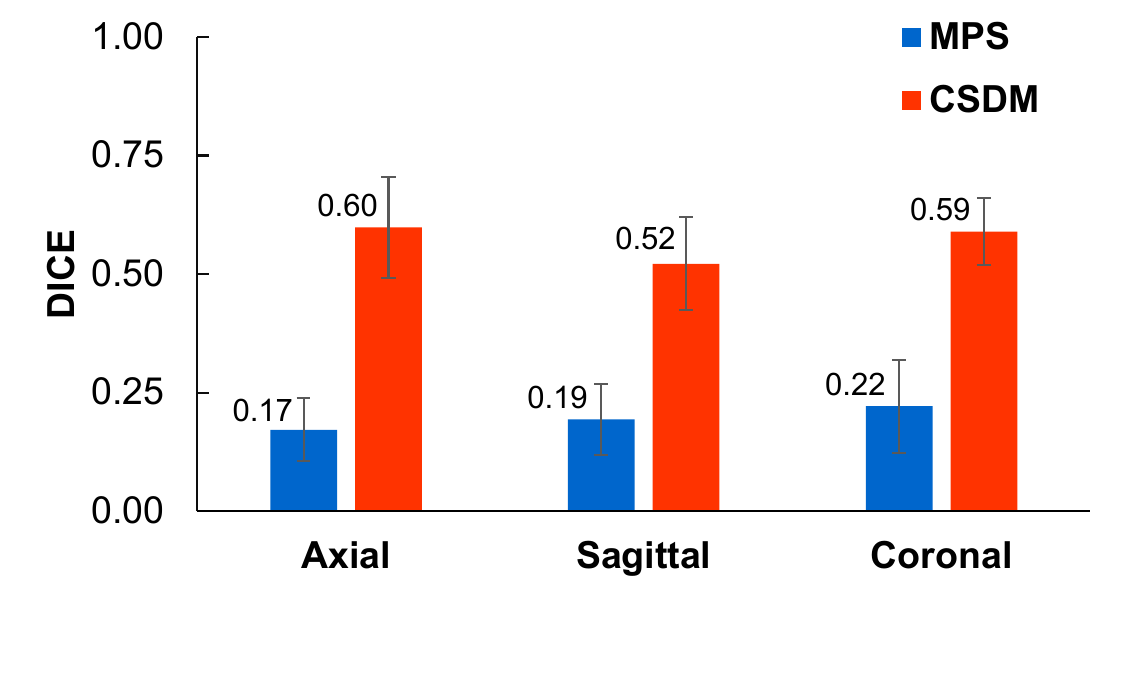}
    \caption{The similarity of the spatial distributions of strain based injury metrics in the gyrencephalic and lissencephalic model is obtained using Sørensen DICE coefficient (DICE). We find lower similarity in the MPS distributions as compared to the CSDM15 distributions under all the directions of rotations. 
    }
    \label{fig:DICE}
\end{figure}

The DICE similarity coefficient (mean ± SD) between gyrencephalic and lissencephalic models for MPS distributions ranged between 0.17 ± 0.07 under the axial rotations, 0.19±0.08 under the sagittal, and 0.22 ± 0.10 under the coronal rotations, highlighting the differences between the gyrencephalic and lissencephalic models. The CSDM15 DICE ranged between 0.60 ± 0.10 under the axial, 0.52 ± 0.10 under the sagittal, and 0.59 ± 0.08 under the coronal loading (Figure \ref{fig:DICE}). The DICE for CSDM15 is higher than for MPS, since CSDM15 DICE essentially discretizes the MPS distribution in steps of 1.0, while MPS DICE considers similarity in steps of 0.025. 

\label{Results:TimeHistories}

\subsection{Effect of cortical folds on peak injury metrics}
\label{Results:TimeHistories}

\begin{figure}[htbp]
    \centering
    \includegraphics[width=1.\textwidth]{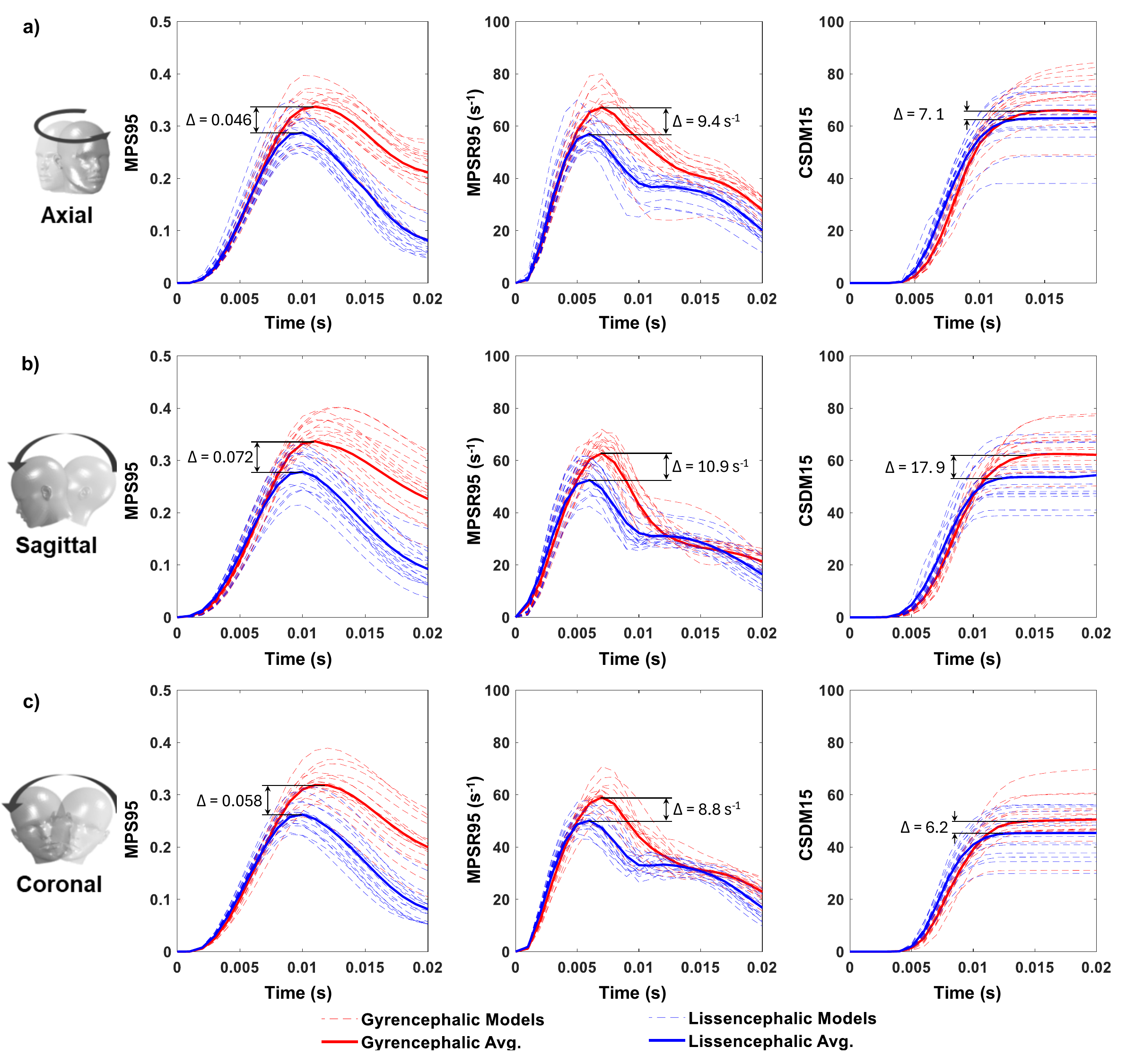}
    \caption{The MPS95, MPSR95, and CSDM time histories from the gyrencephalic (red) and lissencephalic (blue) models for all the subjects under the three rotation directions: a) Axial, b) Sagittal, and c) Coronal. The individual responses are dashed curves and the average response is a solid lines. We see that the lissencephalic model under-predicts all the injury metrics for all the subjects under all loading directions.
    }
    \label{fig:all_time_histories}
\end{figure}

The time history plots of injury metrics, including MPS95, MPSR95, and CSDM, for all the subjects and rotation directions show that the lissencephalic models consistently under-predicted all injury metrics (Figure \ref{fig:all_time_histories}). 
The time to peak MPS95 and MPSR95 is also lower in the lissencephalic models for all loading directions (Figure \ref{fig:all_time_histories}), due to higher strain concentrations occurring deeper in gyrencephalic models, where the shear wave arrives later compared to the cortical surface, which are the highest strain locations in the lissencephalic models. 
This is also reflected in the CSDM15, which is initially higher in the lissencephalic models, before being surpassed by the gyrencephalic models Faster shear wave speed in stiffer brain tissue than the significantly more compliant CSF could also contribute to the differences in CSDM15, which rises faster in the lissencephalic model than in the gyrencephalic model.

\begin{figure}[htbp]
    \centering
    \includegraphics[height=0.4\textwidth]{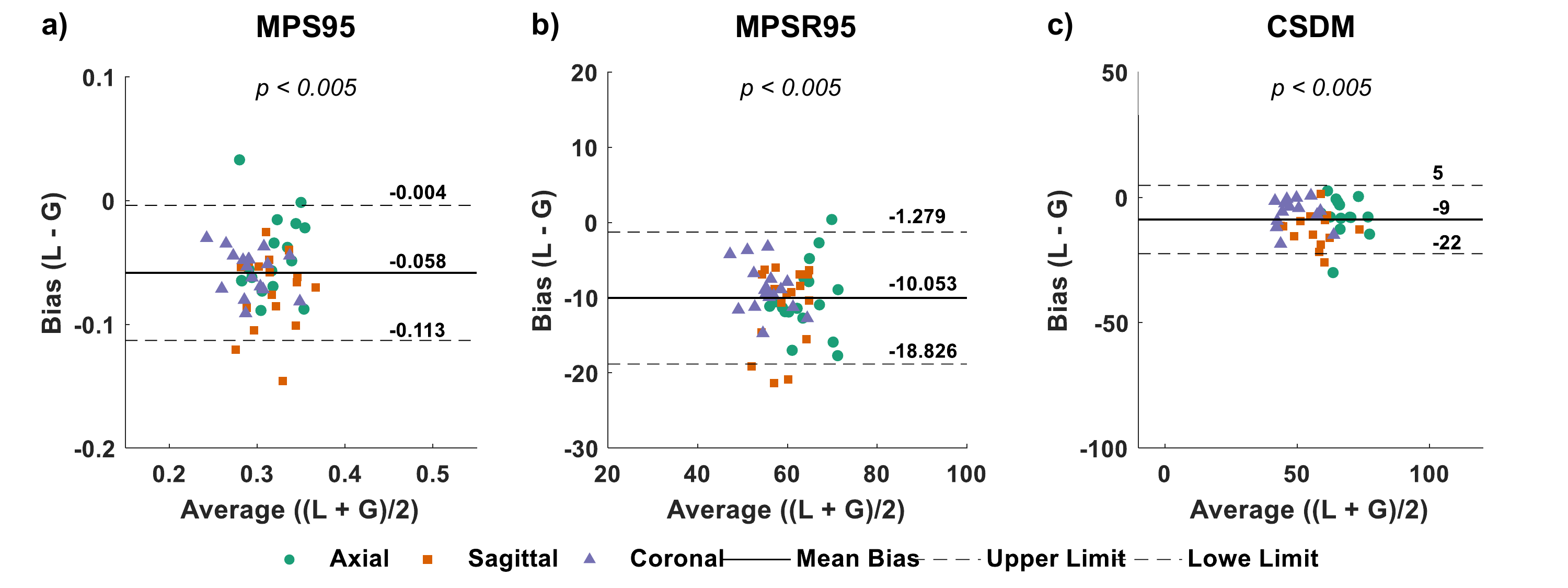}
    \caption{Bland-Altman analysis showing the mean bias and the limits of agreement (mean±1.96×SD) between the gyrencephalic (G) and lissencephalic (L) peak injury metrics: a) MPS95, b) MPSR95, and c) CSDM15. These trends show the gyrencephalic models overpredict the injury metrics compared to the lissencephalic models with high statistical significance.
    }
    \label{fig:Corr_BA}
\end{figure}

The Bland-Altman analysis provided the bias (mean $\pm$ std. dev.) in the peak MPS95 to be $-0.061 \pm 0.026$ (statistically significant, $p<0.005$), in peak MPSR95 to be $-10.05\pm4.48$ s$^-1$ ($p<0.005$), and in CSDM15 to be $-9\pm6.6$ ($p<0.005$) (Figure \ref{fig:Corr_BA}). Normalizing the bias by the maximum injury metric values experienced by the lissencephalic models provides the percentage bias (mean $\pm$ std. dev.) to be $-21.7 \pm 9.1 \%$ in MPS95, $-17.1\pm7.6\% $ in MPSR95, and $-14.4\pm11.3\% $ in CSDM15, highlighting the influence of modeling cortical folds on peak injury metric values. 
No proportional bias was observed in any injury metric.


\subsection{Effect of modeling cortical folds on different regions of interest}
\label{res:Results_ROI}
In this section, we compare the gyrencephalic and lissencephalic simulation results for different regions of interest (ROI) to determine the regional extent of cortical folds' influence. 

\subsubsection{Cerebrum}
The cerebrum is geometrically different between the gyrencephalic and the lissencephalic models. The region experienced overall higher MPS95, MPSR95, and CSDM15 in the gyrencephalic than in the lissencephalic models for all three loading directions.
The Bland-Altman analysis provided the bias (mean $\pm$ std. dev.) in peak MPS95 to be $-0.053\pm0.043$ (statistically significant, $p < 0.005$), in peak MPSR95 to be $-8.53\pm8.56$ s$^{-1}$ ($p < 0.005$), and in CSDM15 to be $-6.6\pm 6.4$ ($p < 0.05$) (Figure \ref{fig:ROI_BA}). 
The bias values translate to $-15.6 \pm 12.7\%$, $ -12.7 \pm 12.8\%$, and $-18.2 \pm 17.5\%$ of the maximum MPS95, MPSR95, and CSDM15 in the gyrencephalic models cerebrum across all models, respectively. No proportional bias was observed in any injury metric.

Since, the cerebral gray and white matter are modeled differently in the gyrencephalic and lissencephalic models (Figure \ref{Fig:workflow}a), the size of high-strain regions on the cortical surface can artificially cause differences in the peak injury metrics in these regions. For example, a point at a depth of 6 mm from the cortical surface will lie in the white matter in the lissencephalic model, but due to the presence of sulci 2 mm from it, it will fall in the gray matter in the gyrencephalic model. Therefore, we don’t study the gray and white matter separately, but combined.

\subsubsection{Deep Brain Regions}

The deep brain regions (Section \ref{sec:FEMdevelopment}) is geometrically identical between the gyrencephalic and the lissencephalic models. However, the distance between the cortical folds and the deep brain regions can be as low as $\sim$10 mm between deep sulci, such as the Sylvian fissure, or medial Cingulate sulci and outer regions of the putamen or caudate, as opposed to at least $\sim$30 mm from the cortical surface in the lissencephalic models. The region experienced overall higher strain in the gyrencephalic models than the lissencephalic ones for all three loading directions ($p < 0.005$), arising from the zone of strain concentrations at the base of the sulci.

We found the mean bias ($\pm$ std. dev.) in peak MPS95 was found to be $-0.057\pm0.038$, in peak MPSR95 to be $-12.88\pm7.28$ s$^{-1}$, and in CSDM15 to be $-17.1\pm12.1$ (all $p<0.005$) (Figure \ref{fig:ROI_BA}). The normalized bias values are  $-25.3 \pm 17.0 \%$, $-28.9 \pm 16.3 \%$ and  $-46.8 \pm 33.3 \%$,  in MPS95, MPSR95, and CSDM15, respectively. 
A proportional bias was observed again in all injury metrics, with the bias increasing with the magnitude of the injury metric.

\subsubsection{Corpus Callosum}

The corpus callosum is geometrically the same between the gyrencephalic and the lissencephalic models, although is located close to the gyri on the medial axis. It experienced one of the highest peak values for all injury metrics and also the highest biases between the two models across all loading directions, since it sits adjacent to the lateral ventricles. 
The high strain in the brain tissue around the ventricles in the gyrencephalic model is passed on to the ventricles, which also experience higher strains than in the lissencephalic model. The tissues near the ventricles are stretched higher as the wave passes through the ventricles to the other side of the brain, eventually affecting the regions far removed from the cortical folds, including the corpus callosum. It experienced higher strain in the gyrencephalic than in the lissencephalic models for all three loading directions ($p < 0.005$).

The Bland-Altman analysis provided the bias (mean $\pm$ std. dev.) in peak MPS95 to be $-0.178 \pm 0.143$, in peak MPSR95 to be $-38.05\pm32.65$ s$^-1$, and in CSDM15 to be $-19.1 \pm 19.7$ (all $p<0.005$). (Figure \ref{fig:ROI_BA}). The normalized bias values are $-48.6 \pm 39.0 \%$, $-49.4 \pm 42.4 \%$ and  $-52.3 \pm 53.8 \%$, in MPS95, MPSR95, and CSDM15, respectively. 
We observe a proportional bias in MPS95 and MPSR95, with the bias increasing with magnitude in both injury metrics. The CSDM15 shows proportional bias under the axial and coronal loading directions, but not under the sagittal direction.

\subsubsection{Cerebellum}

The cerebellum is geometrically identical between the gyrencephalic and the lissencephalic models, and away from the cortical folds, separated by tentorium. The region experienced overall higher strain in the gyrencephalic models than the lissencephalic ones for all three loading directions, due to its proximity to the fourth ventricles. 

The bias (mean $\pm$ std. dev.) in peak MPS95 was found to be $-0.035\pm0.042$ ($p<0.005$), in peak MPSR95 to be $-7.18\pm6.51$ s$^{-1}$ ($p<0.05$), and in CSDM15 to be $-10.8\pm11.5$ ($p<0.05$) (Figure \ref{fig:ROI_BA}). The normalized bias values are  $-16.4 \pm 19.9 \%$, $-16.2 \pm 14.7 \%$ and  $-29.6 \pm 31.6 \%$, in MPS95, MPSR95, and CSDM15, respectively. 
We observe proportional bias in all injury metrics, with the bias increasing with the magnitude of the injury metric.

Given the resolution of our model (1 mm), all the folds in the cerebellum are also not fully resolved, and this study keeps the cerebellum identical between the two models, and only investigates the effects of folds in the cerebrum.

\subsubsection{Brain Stem}

The brain stem is also geometrically identical between the gyrencephalic and the lissencephalic models and separated from the cortical folds, and experienced higher peak values for all injury metrics compared to other ROIs, except the corpus callosum.
The region experienced overall higher strain in the gyrencephalic models than the lissencephalic ones for all three loading directions ($\textit{p}<0.005$)due to its proximity to the fourth ventricles.

\begin{figure}[htbp]
    \centering
    \includegraphics[width=0.98\textwidth]{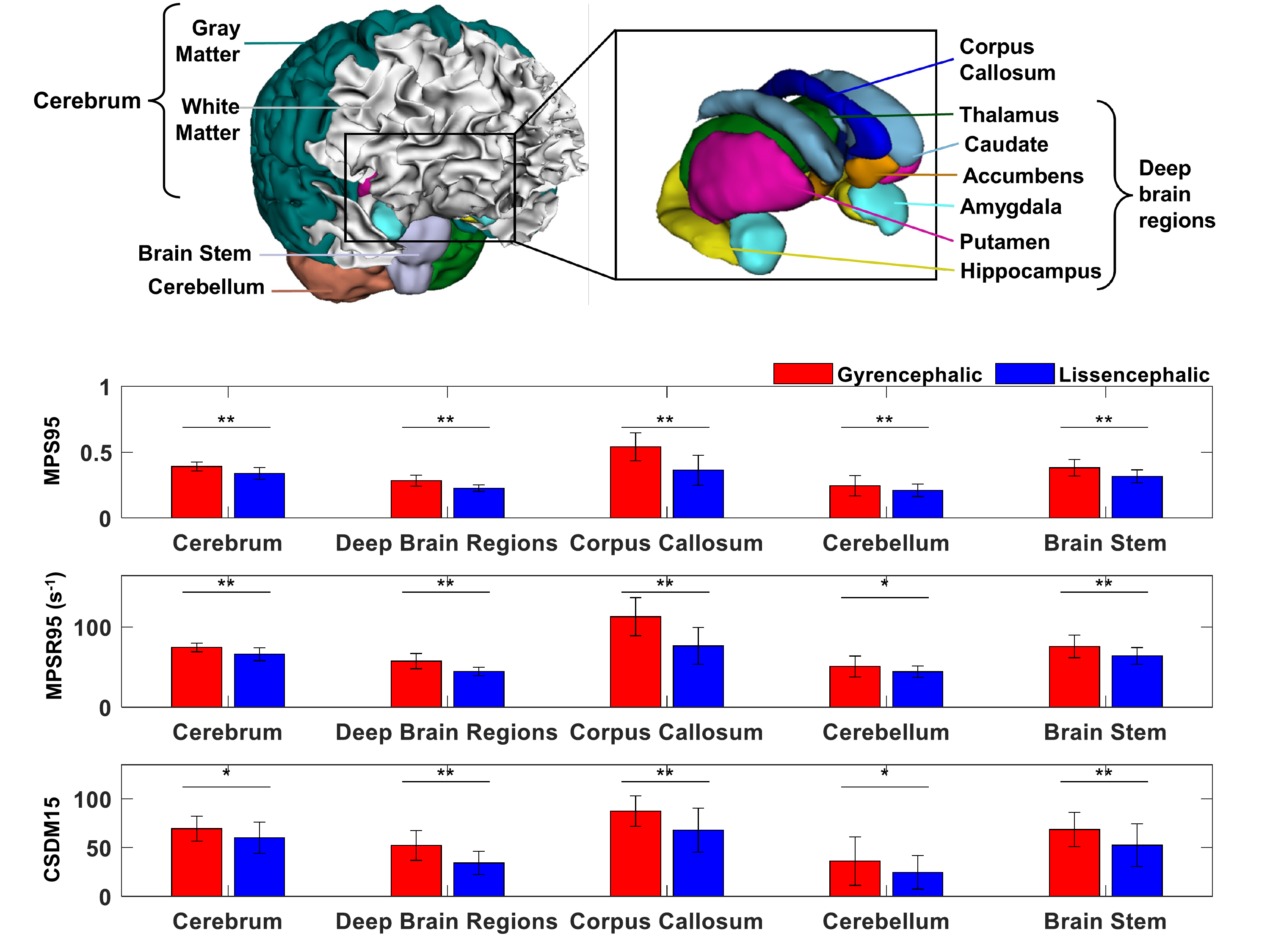}
    \caption{Different regions of interest (ROI) showing the cerebrum, cerebellum, deep brain regions, corpus callosum, and brain stem. The statistical significance of the differences is marked for the peak injury metrics MPS95, MPSR95, and CSDM15. 
    These trends show that the gyrencephalic models predict higher peak injury metrics compared to the lissencephalic models with high statistical significance in all ROIs, except for the cerebellum, which sits far removed from the cortical folds and separated by the tentorium.
    }
    \label{fig:ROI_BA}
\end{figure}

The bias (mean $\pm$ std. dev.) in peak MPS95 was found to be $-0.065\pm0.050$, in peak MPSR95 to be $-11.17\pm10.85$ $s^{-1}$, and in CSDM15 to be $-13.0\pm13.7$ (all $p<0.05$) (Figure \ref{fig:ROI_BA}). The normalized bias values are $-20.5 \pm 15.6 \%$, $-17.4 \pm 16.9 \%$ and  $-35.5 \pm 37.5 \%$,  in MPS95, MPSR95, and CSDM15, respectively. 
We found no proportional bias in any injury metric.

These results show that modeling the cortical folds affects the injury metrics even in the ROIs that have not been altered between the two models, namely the corpus callosum, brain stem, and deep brain regions.

\section{Discussion}

In this study, we aimed to determine the need to incorporate cortical folds in FE head models for accurately predicting the risk of mTBI under head rotational acceleration events. We attempted to answer this question by comparing different mTBI injury metrics between the gyrencephalic and lissencephalic FE models of 18 subjects aged 9 - 18 years. The models were used to simulate an idealized concussive event with the rotational acceleration of 10 krad/s$^2$ and rotational velocity of 60 rad/s about each principal axis. In this section, we discuss our results in comparison with other studies to understand the factors affecting the importance of modeling cortical folds (Section \ref{Diss:comp}) and the limitations of our study and future works (Section \ref{Diss:lim}).

\subsection{Comparison with other studies}\label{Diss:comp}

Several previous studies have investigated the role of cortical folds on mTBI injury metrics under head acceleration events and have found conflicting results.

Some studies are based on the full brain models, while some are based on mesoscale models of the region incorporating approximately a couple of gyri.
Among the studies at the full brain scale, some predicted higher injury metrics in gyrencephalic models \cite{mazurkiewicz2021impact,fagan2020simulation}, some in lissencephalic models \cite{ho2009can}, and some found different trends in different metrics \cite{song2015finite}.
The studies varied in terms of the applied loading conditions, the analyzed injury metrics, and the material properties used in the models. While the studies focused on impact loading, most studies only incorporated translational accelerations \cite{fagan2020simulation,saez2020topological,song2015finite,mazurkiewicz2021impact}. It has been shown that during an impact, the resulting translational acceleration leads primarily to high pressure, while the angular kinematics primarily leads to the shear strains in the brain tissue \cite{zhang2006role}. The low pressure from translational acceleration is correlated to cavitation in the brain tissue, while shear strain is directly related to neuronal stretch and death. Since an impact loading results in negative pressures much lower than cavitation thresholds, the rotational acceleration of the head is more critical in an impact loading \cite{tripathi2025laboratory,carlsen2021quantitative}. Therefore, our study focuses on rotational head accelerations, and we can only compare our results with studies that incorporate angular acceleration. 

The low impact of translational acceleration on brain tissue is reflected in the low strains reported in these studies ($<5\%$). Fagan et al. \cite{fagan2020simulation} found higher strains in gyrencephalic models but similar pressure response in the two models; Mazurkiewicz et al. \cite{mazurkiewicz2021impact} also found higher strains in gyrencephalic model; Saez et al. \cite{saez2020topological} found similar strains in the two models but lower pressures and maximum principal stresses in gyrencephalic model; and Song et al. \cite{song2015finite} found no difference in the pressure, higher stress and lower strains in gyrencephalic models. The differences in these results could stem from the differences in the models used. 
Saez et al. modeled a 3D models of a human (high gyrification), a macaque (lower gyrification), and a mouse (lissencephalic) scaled to the same size, Fagan et al. developed a 2D gyrencephalic and lissencephalic models of human axial slice, Mazurkiewicz et al. conducted experiments on a 2D gyrencephalic and lissencephalic surrogates based on coronal slice of porcine brain, and Song et al. studied the sagittal slice of a human head. The studies also differed in terms of material properties, e.g. Fagan et al. modeled the CSF as an Eulerian mesh, resulting in separation between the brain and the CSF; while the CSF material modeling was not reported in the Saez et al. paper. These differences make the direct comparison difficult to identify the factors affecting the role of cortical folds.


Ho and Kleiven \cite{ho2009can} investigated the effect of cortical folds by comparing strains from the 3D lissencephalic and gyrencephalic models of a human subject under translational and rotational loading conditions. They found that lissencephalic models experience higher strains under a 5 ms, 10 krad/s$^2$ loading, in contrast to our findings under similar loading. 
The opposite trends arise from the differences in the skull-brain interface in the head models used in the studies. Ho and Kleiven modeled the CSF as an elastic fluid, and also incorporated the Pia mater (shear stiffness ~10 MPa) as a viscoelastic shell, which significantly increased the skull-brain interface stiffness. 
Since the Pia mater is more than 1000$\times$ stiffer than the brain tissue, this region in the skull-brain interface becomes stiffer than the brain material, as opposed to the model in our study, where the skull-brain interface is softer than the brain tissue.

Since incorporating the Pia mater in a voxel mesh with a jagged interface was not feasible in our model, we approximately studied the effect of a stiffer skull-brain interface by modeling the sub-arachnoid space with a higher stiffness of 20kPa, which is stiffer than the long-term modulus of the brain (Supplementary Section S3). Another FE head model has used a similarly stiff sub-arachnoid material to account for the trabeculae and other connective tissues in the region \cite{ghajari2017computational}. The result of this preliminary analysis shows that as the CSF stiffness increases, the strain experienced by the brain tissue decreases significantly in the gyrencephalic model. The CSF deformation decreases with the CSF stiffness, and therefore, the strain wave transmitted to the brain is lower than in the case of a softer CSF. 
Since, the sulci in the gyrencephalic models are filled with CSF, the increased stiffness of the skull-brain interface reduces the strain in the gyrencephalic model more significantly as compared to the lissencephalic model, which experiences a smaller decrease in the strain, resulting in higher strains in the lissencephalic model compared to the gyrencephalic one. Unfortunately, there is no consensus on the best approach to simulate the skull–brain interface, and we the choice of skull-brain interface in our model is based on previously validated models \cite{carlsen2021quantitative,nakarmi2024role,fagan2020simulation,giudice2021calibration,mao2013development}.

The other important distinction with the Ho and Kleiven study is the injury metric analyzed. The study looked into the 100th percentile strain, which doesn’t remove the effect of numerical artifacts, as compared to the 95th percentile used in our study. It is also worth mentioning that the quality of segmentation that captures the depth of sulci in the model can be affected by the choice of segmentation tools, which can also contribute to different trends. 

The finding of our study that the gyrencephalic models experience higher peak injury metrics
is corroborated by most mesoscale models-based studies  \cite{bakhtiarydavijani2021mesoscale,he2022mesoscale,cloots2008biomechanics}. 
However, the boundary conditions to capture the impact loading in these studies are not consistent. Most studies apply compression loading to compare the pressure or stress responses \cite{saboori2012effect,he2022mesoscale,bakhtiarydavijani2021mesoscale}. The studies found increasing stress with increasing depth of the sulci. While \cite{bakhtiarydavijani2021mesoscale} also analyzed a shear loading case, they only reported stresses. 
Therefore, a direct comparison with these studies are not feasible. 
On the other hand, Cloots et al. \cite{cloots2008biomechanics} approximated the impact event using a shear-acceleration profile in the region of interest, derived from the output of a full-head model simulation under linear acceleration. They also found higher strains and stresses in models with sulci than in a homogeneous model and different strains for different sulci geometries, similar to the results of our study. However, since the model only captures sulci, its interaction with the rest of the brain, especially the ventricles can’t be captured in this model. 
On the other hand, since the mesh resolution in our FE models is not as fine as in the meso-scale models, we are unable to capture the strain localization at the base of the sulci. 

Comparing the results of our study with the existing literature, we find that modeling cortical folds affects the spatial distribution and peak values of the strain-based injury metrics under head rotations. However, the magnitude of differences due to modeling cortical folds is affected by the choice of skull-brain interface and brain stiffness; the mode of loading, translational or rotational; and the injury metrics analyzed.

\subsection{Limitations and Future Works}\label{Diss:lim}

There are several limitations to our study. First, we only analyze a single head acceleration magnitude of 10 krad/s$^2$, which lies towards the higher end of the range of measured impacts. The duration of the loading (10 ms) also synchronizes with the head natural frequency to produce higher strains than other duration \cite{carlsen2021quantitative}. While we anticipate smaller differences in the peak strains between the lissencephalic and gyrencephalic models as the magnitude of peak strains decreases with other loading conditions, 
given the complex non-linear relationship between loading and peak injury metrics  \cite{carlsen2021quantitative}, further analysis is required to access the importance of modeling cortical folds in a different range of loading conditions. 
However, the general results of this study will still apply to the strain distributions affected by the cortical folds.

Our model didn't distinctly model the Pia mater, the vasculature in the sub-arachnoid space, and the CSF in the head is also not modeled as a fluid. We modeled the space as a very soft solid to account for all the structures. Unfortunately, the best approach to simulate the skull-brain interface remains not unresolved, but will greatly affect the importance of modeling the cortical folds. Future studies should investigate the effect of modeling Pia mater and cerebral vasculature on the strains experienced by the brain.  

We selected a voxel mesh for the FE model, since a conformal mesh requires significantly higher development time and manual intervention, prohibiting scaling to multiple subject-specific models. The use of mesh morphing based on existing image registration methods provided limited subject-specificity (DICE $< 0.7$) near the gray matter-CSF interface \cite{li2021anatomically}. Since recent studies report small differences between a voxel model and a conformal model under a rotational loading \cite{zhou2025surface}, the voxel mesh was a reasonable choice for this study.


Given the resolution of our model, if two gyri touch each other, they share common nodes and therefore are not free to detach from each other. Future studies should investigate the effect of sliding between the gyri. Our FE models also lacked mesh adaptivity or high mesh density near sharp corners, such as cortical sulci and ventricles, limiting the ability to capture the strain concentrations at the sulci. These limits the ability of our model to capture tau-protein accumulation patterns seen in CTE \cite{mckee2009chronic}. Future studies should move towards higher mesh density near high curvature points.

Despite the limitations in our workflow, the results of the study show a significant effect of cortical folds on multiple injury metrics, across multiple subjects, and under different loading directions, highlighting the need to incorporate cortical folds in future studies on mTBI risk assessment of different activities.

\backmatter

\bmhead{Acknowledgments}
The authors gratefully acknowledge the PANTHER program for facilitating fruitful discussions and collaborations.

\section*{Declarations}

\begin{itemize}
\item \textbf{Funding:} The authors gratefully acknowledge the support from the University of Wisconsin-Madison Office of the Vice Chancellor for Research (OVCR) and the Athletic Department. Funding for this award has been provided through Big 10 Athletic Media Revenue (136-AAI3375). The authors also acknowledge the U.S. Office of Naval Research funding under the PANTHER award N00014-21-1-2044 through Dr. Timothy Bentley.

\item \textbf{Competing interests:} The authors have no competing interests to declare that are relevant to the content of this article. 

\item \textbf{Author Contributions:} All authors contributed to the study conception and design.  The FE modeling, simulations, post-processing, and data analyses were performed by Anu Tripathi. The medical images were acquired by Peter Ferrazzano. The original draft was prepared by Anu Tripathi and revised by all authors. All authors reviewed and approved the final manuscript.

\item \textbf{Supplementary Information} The supplementary material is available \href{run:Supp.pdf}{here}.
\end{itemize}

\noindent

\bigskip



\bibliography{sn-bibliography}

\end{document}